\documentclass[twocolumn,showpacs,preprintnumbers,amsmath,amssymb]{revtex4-1}
\usepackage[dvipdfmx]{graphicx}
\usepackage{dcolumn}% Align table columns on decimal point
\usepackage{bm}% bold math

\newcommand{\be}{\begin{eqnarray}}
\newcommand{\ee}{\end{eqnarray}}

\begin{document}

%\preprint{APS/123-QED}

\title{Phase diagrams of Bose-Hubbard model and Haldane-Bose-Hubbard model
with complex hopping amplitudes}
% Force line breaks with \\

\author{Yoshihito Kuno, Takashi Nakafuji, and Ikuo Ichinose}
% \altaffiliation[Also at ]{}%Lines break automatically or can be forced with \\
\affiliation{
$^1$Department of Applied Physics, Nagoya Institute of Technology, 
Nagoya  466-8555, Japan}

\date{\today}% It is always \today, today,

\begin{abstract}
In this paper, we study Bose-Hubbard models on the square and honeycomb
lattices with complex hopping amplitudes, which are feasible by recent experiments
of cold atomic gases in optical lattices. 
To clarify phase diagrams, we use an extended quantum Monte-Carlo 
simulations (eQMC).
For the system on the square lattice, the complex hopping is realized by
an artificial magnetic field.
We found that vortex-solid states form for certain set of
magnetic field, i.e., the magnetic field with the flux quanta per plaquette 
$f=p/q$, where $p$ and $q$ are co-prime natural numbers.
For the system on the honeycomb lattice, we add the next-nearest neighbor
complex hopping.
The model is a bosonic analog of the Haldane-Hubbard model.
By means of the eQMC, we study the model with both weak and
strong on-site repulsions.
Numerical study shows that the model has a rich phase diagram.
We also found that in the system defined on the honeycomb lattice of 
the cylinder geometry, an interesting edge state appears.
\end{abstract}

\maketitle
%%%%%%%%%%%%%%%%%%%%%%%%%%%%%%%%%%%%%%%%%%%%%%%%%%%%%%%%%%%%%%%%%

\section{Introduction} \label{intro}

Cold atoms in an optical lattice (OL) have been used as versatile quantum simulators
for the last decade.
In particular, the Bose gas system is highly controllable and has a
rich phase diagram as a strongly-correlation system\cite{Bloch}. 
Recently, generation of an artificial magnetic field in cold atom systems was
experimentally succeeded by using laser-assisted tunneling in a tilted optical 
potential\cite{staggeredmg,uniformmg},
whose theoretical proposal was given by Jaksch and Zoller\cite{Zoller1}. 
These experimental methods can create a stronger magnetic field 
compared to that generated by rotating optical lattice\cite{rotatingmg},
and therefore it is expected that a strong-magnetic field regime corresponding to,
e.g., the quantum Hall state is realized in cold atom systems.

Bose gas system in a strong magnetic field has been studied very actively in the 
last several years.
In particular, study on two-component Bose gas system in a strong magnetic field
in the continuum space predicts a bosonic analog of integer quantum Hall
state (IQHS) for certain inter and intra-repulsions\cite{Furukawa}.
Generation of an incompressible vortex liquid was also suggested
in a similar parameter region of Bose system in a strong magnetic field\cite{Cooper}.
On the other hand for the Bose gas system on the optical lattice, 
the numerical exact diagnalization suggests an existence of a bosonic Laughlin 
state\cite{Sorensen,Hafezi}, and also a new kind of fractional quantum 
Hall state\cite{Cooper2}.
 
There are many studies on the lattice boson systems with a dilute particle density 
and in a weak magnetic field\cite{Bhat,Lundh}.
Formation of the vortex solid was predicted there.
However for the system of interacting bosons in a strong magnetic field,
the structure of the ground-state, in particular, structure of all the vortex-solid states 
are not known. 
Study on the complete global phase diagram of the system for various hopping amplitude, 
interactions, magnetic-field strength and particle filling, etc., is still missing.
For lattice boson systems and certain related models in strong magnetic regime, 
a number of works have been reported so far. 
For classical spin models, Choi and Doniach\cite{Doniach}
found by an analytical method that the uniformly frustrated two-dimensional (2D) 
XY model has stable vortex-solid ground-states at $f=1/2$, $1/3$ and $1/4$, 
where $f$ is the magnetic flux quanta per the fundamental plaquette.
Also, some of vortex-solid states were predicted by using Monte-Carlo 
simulations\cite{Nakano, KKLee,KasamatsuMC} and a Gross-Pitaevskii 
theory\cite{Kato,KasamatsuGP};
these studies showed some vortex-solid patterns for the 2D classical model at
$f=1/2$, $1/3$ and $1/4$. 
Our previous study by using effective theory and Monte-Carlo simulations also 
predicted some solid patterns at $f=1/2$, $1/3$ in two component lattice boson 
system assuming commensurate particle density\cite{KKK0}.   

In theoretical models describing systems in a magnetic field, the hopping 
amplitudes acquire a nontrivial phase and become complex numbers.
As a result, there appear various interesting phases, some of which are aforementioned
vortex solid, the bosonic Laughing state, etc. 
In the first half of the present paper, we clarify various vortex ground-states 
of a Bose Hubbard model (BHM) in a strong magnetic field by using 
extended quantum Monte-Carlo simulations (eQMC), in which effects of 
both phase and density fluctuations of the boson field are taken into account properly
in the path-integral formalism.
We found that a phase transition from the vortex solid to vortex liquid
takes place as the on-site repulsion $U$ is varied. 
In the second half of the paper, we study a bosonic analog of the Haldane
model on the honeycomb lattice with the on-site repulsion $U$, which is sometimes
called Haldane-Bose-Hubbard model (HBHM).
Besides the nearest-neighbor (NN) hopping, the HBHM contains the next-NN (NNN)
hopping with a nontrivial phase $\phi$.
The Haldane model was originally proposed as a fermion model that has 
the ground-state similar to the quantum Hall state\cite{Haldane}.
In the recent paper\cite{Hofstetter}, the HBHM with $\phi=\pi/2$ 
was studied by the dynamical mean-field theory and an exact
diagonalization of a small system.
In this paper, we study the HBHM by the eQMC for both strong and weak on-site 
repulsion.
We exhibit the global phase diagram, which should be compared with that obtained
in Ref.\cite{Hofstetter}, detailed critical behaviors of the phase
transitions, physical properties of an edge state in a cylinder geometry, etc.

This paper is organized as follows.
In Sec.II, we introduce the BHM in an artificial magnetic field and 
explain the derivation of the effective theory, which will be studied by the eQMC. 
In Sec.III, we explain some details of the eQMC show the global phase diagram, 
which is obtained by the eQMC.
In the phase diagram, there exist various vortex-solid states and also
vortex quantum liquid states for certain specific magnitude of the magnetic field. 
Phase transition from the disordered state to vortex solid is studied by 
the finite size scaling and the critical exponents are estimated.
In Sec.IV, the BHM in an magnetic field is studied by using a duality transformation
and origin of the vortex-solid patterns found in Sec.III is explained.
Phase diagram obtained by varying boson density is also shown.
From these observation, possibility of bosonic Laughlin state is discussed
by using a Chern-Simons theory.
In Sec.V, the phase diagram of the HBHM is obtained, which has four phases.
Detailed study of each phase is given by calculating the expectation value of the
current and phase correlation on links.
Finally we investigate the HBHM in a cylinder geometry, in particular, we are
interested in the edge state.
Sec.VI is devoted for conclusion.

%%%%%%%%%%%%%%%%%%%%%%%%%%%%%%%%%%%%%%%%%%%%%%%%%%%%
\section{Bose Hubbard model in a uniform magnetic field and the effective model}\label{Model}

In this section, we consider the BHM defined on 
a two dimensional (2D) square lattice. 
We start with the BH Hamiltonian in an artificial magnetic field with a
vector potential $A_{\mu}(r)$, 
\begin{eqnarray}
H_{\rm BH}=&&-J\sum_{r,\mu} a^{\dagger}_{r}e^{-iA_{\mu}(r)} a_{r+\mu}
+\mbox{h.c.}  \nonumber   \\
&&+\sum_{r}U n^{2}_{r},
\label{BH}
\end{eqnarray}
where $r$ denotes sites of the lattice and $\mu=\hat{x},\hat{y}$ is the direction 
index and it sometimes denotes the unit vector.
$a_{r}(a^{\dagger}_{r})$ is a bosonic annihilation (creation) operator and
$n_{r}= a^{\dagger}_{r} a_{r}$. $J$ is hopping amplitude and $U$ is on-site repulsive interaction. 
All these parameters are highly controllable in experiments\cite{Bloch}. 
$A_{\mu}(r)$ represents a uniform magnetic field perpendicular to the lattice plane.
In the numerical calculation, we use
\begin{eqnarray}
A_{\mu}(r)=\left\{ \begin{array}{ll}
-2\pi f y & (\mu=\hat{x}) \\
0 & (\mu=\hat{y}) \\
\end{array}\right. ,
%\label{U(1)gauge}
\end{eqnarray}
where $r=(x,y)$ and $f$ is the magnitude of the magnetic flux per plaquette, 
and its range is $0 \leq f \leq 1$ due to compactness of 
the phase degrees of freedom.
The gauge field $A_{\mu}(r)$ creates the uniform magnetic field $B_{z}$; 
$B_{z}=\sum_{p}A_{p}=2\pi f$, where $p$ denotes a directed close path around 
a plaquette and $A_{p}$ is the vector potential on the path.
This model is experimentally feasible\cite{uniformmg}.

In the following sections, we shall study the BHM in Eq.(\ref{BH})
by the numerical MC simulations.
To this end, we have to derive an effective model for the BHM with
{\em a positive definite action}. 
Although it is difficult to perform the direct quantum simulations  
because of the complex hopping in Eq.(\ref{BH}), 
we can derive a useful effective model including relevant quantum effects 
by integrating out certain degrees of freedom in the path-integral formalism.
 
In previous work\cite{KKK0}, we derived the effective model for some related bosonic 
systems, and the effective model obtained there was numerically 
studied by the MC simulations. 
In this paper, we shall extend the previous methods in order to search inhomogeneous
states as a ground state. 

Let us start the derivation of the effective model mentioned above.
We first parameterize the boson variables in the path integral as 
$a_{r}=\sqrt{\rho_{r}+\delta \rho_{r}}\ e^{i\theta(r)}$, where
$\rho_{r}$ is the mean density at site $r$, which
is regarded as {\em variational parameter} in the MC simulation (see later discussion). 
On the other hand, 
the variable $\delta \rho_{r}$ represents {\em quantum fluctuation} of the
density around the mean value $\rho_r$, and then the Berry phase term in the
action is given as $(\partial_{\tau}\theta(r))\delta\rho_{r}$, where
the variable $\theta(r)$ is the phase of the boson field. 
By substituting the above parameterization, the partition function of 
the BHM defined by Eq.(\ref{BH}) is given as follows by using the imaginary-time $\tau$,
\begin{eqnarray}
Z_{\rm BH}&=&\int [d\delta \rho_{r}][d\theta(r)]e^{-S_{\rm BH}},\label{ZqXY1}\\
S_{\rm BH}&=&\int d\tau \biggr(\sum_{r}i(\partial_{\tau}\theta(r))\delta\rho_{r}\nonumber\\
&-&\sum_{r,\mu}J\sqrt{\rho_{r}\rho_{r+\mu}}\cos(\theta(r)-\theta(r+\mu)+A_{\mu}(r))\nonumber\\
&\hspace{-0.5cm}+&\hspace{-0.4cm}J{\cal T}(\rho,\theta)\delta\rho+
\sum_{r}(U\rho^{2}_{r}+U\delta\rho^{2}_{r}+2U\rho_r\delta\rho_r) \biggl),
\label{SBH2}
\end{eqnarray}
where $J{\cal T}(\rho,\theta)\delta\rho$ denotes the first-order contribution of the quantum fluctuation $\{\delta\rho_r\}$ in the hopping term $J$, and we have
neglected the higher-order terms of $\{\delta\rho_r\}$ coming from the $J$-terms.
As explain in the following section, we determine $\{\rho_r\}$ by 
the minimum-energy condition in the MC calculation.
Therefore the terms of $O(\delta\rho)$ in Eq.(\ref{SBH2}), except the Berry phase,
cancel with each other.
Then in Eq.(\ref{ZqXY1}), we perform the Gaussian integral for the density fluctuation
$\delta\rho_{r}$ to derive the effective theory whose action is denotes by $S_{\rm qXY}$.
Henceforth, we call this effective model $S_{\rm qXY}$ the 
{\em quantum XY model} (qXYM), whose partition function is given by
\begin{eqnarray}
Z_{\rm qXY} %&=&\int [d\delta \rho_{r}][d\theta(r)]e^{-S_{\rm BH}}\nonumber\\
 &=& \int[d\theta (r)]e^{-S_{\rm qXY}},\\
\label{ZqXY}
S_{\rm qXY}&=&\int d\tau 
\biggr(\sum_{r}\frac{1}{4U}(\partial_{\tau}\theta (r))^{2}\nonumber\\
&-&J\sum_{r,\mu}\sqrt{\rho_{r}\rho_{r+\mu}}\cos(\theta(r)-\theta(r+\mu)+A_{\mu}(r))\nonumber\\
&+&\sum_{r}U\rho^{2}_{r} \biggl).
\label{SqXY}
\label{qXY}
\end{eqnarray}
Here it should be remarked that $S_{\rm qXY}$ in Eq.(\ref{SqXY}) is real,
and then the standard MC simulation is applicable for the numerical study of the system.
Moreover it should be noticed that this model has the {\em Lorentz symmetry} and 
therefore it is expected that an excitation corresponding to the {\em Higgs mode} 
exists\cite{Higgs_theory,KKK0}. 
In fact this mode has been observed in optical lattice experiments\cite{Higgs}.

%%%%%%%%%%%%%%%%%%%%%%%%%%%%%%%%%%%%%%%%%%%%%%%%%%%%%
\section{Extended Quantum Monte-Carlo simulation}\label{EQMC}

In this section, we numerically study the qXYM by means of the eQMC
explained in the previous section. 
In particular, we are interested in vortex dynamics induced by the strong artificial
magnetic field with the vector potential $A_\mu(r)$, and obtain 
a global phase  diagram of vortex states. 
For the study on phase diagram of  
the present system with the complex hopping amplitudes, the qXYM is quite 
useful as the phase degrees of freedom of the boson field plays an essentially
important role.

For the eQMC, we put the lattice spacing of the OL, $a_{L}$, 
to the unit of length and also introduce a discretized lattice for the 
imaginary-time $\tau$ with the lattice spacing $\Delta \tau$.
Thus, the qXYM becomes a kind of 3D XY model defined on the
space-time lattice, whereas its coefficients depend on the variational parameters
$\{\rho_r \}$. 
The lattice action of the qXYM is given as follows: 
\begin{eqnarray}
S^{\rm L}_{\rm qXY}&=&\sum_{r}-\frac{1}{4U\Delta\tau}\cos(\theta (r)-\theta(r-\hat{\tau}))\nonumber\\
&-&J\Delta\tau
\sum_{r,\mu}\sqrt{\rho_{r}\rho_{r+\mu}}\cos(\theta(r)-\theta(r+\mu)
+A_{\mu}(r))\nonumber\\
&+&\sum_{r}U\Delta\tau\rho^{2}_{r},  \nonumber  \\
&\equiv&\sum_{r}-\frac{1}{4U\Delta\tau}\cos(\theta (r)-\theta(r-\hat{\tau}))
+H_{\rm qXY}, 
\label{lattice_qXY}
\end{eqnarray}
where $r$ denotes sites on the 3D lattice.

Here we briefly explain the eQMC to study the qXYM of Eq.(\ref{lattice_qXY}).
The qXYM action on path integral includes both the variational
parameters $\{\rho_{r}\}$ and the dynamical phase variables $\theta(r)$'s. 
We determines the variational variables $\{\rho_{r}\}$ by the
minimum-energy condition by using MC methods. 
More precisely, we express the extended partition function 
$[Z^{\rm L}_{\rm qXY}]$ 
of Eq. (\ref{lattice_qXY}) as  
\begin{eqnarray}
[Z^{\rm L}_{\rm qXY}] &\equiv&  \int [d\rho_{r}]
Z^{\rm L}_{\rm qXY}(\{\rho_{r}\}), \nonumber \\
Z^{\rm L}_{\rm qXY}(\{\rho_{r}\})  &=& \int[d\theta (r)]e^{-S^{\rm L}_{\rm qXY}}.
\label{ZLqXY}
\end{eqnarray} 
In the practical calculation of Eq.(\ref{ZLqXY}), we treat $\{\rho_{r}\}$ as slow
variables in the MC local-update with keeping the mean value of $\{\rho_{r}\}$
constant.
On the other hand, we perform the MC simulation for the dynamical
variables $\theta(r)$ as the ordinary variables for the fixed $\{\rho_{r}\}$.
From the experience, e.g. in Ref.\cite{KKK2}, 
we know that $\{\rho_{r}\}$ are quite stable under local updates 
for given values of parameters in the qXYM action.
In the present study, we verified this stable behavior of $\{\rho_{r}\}$ for 
typical configurations of $\theta(r)$, i.e., once $\{\rho_{r}\}$ is selected for 
fixed parameters of $S^{\rm L}_{\rm qXY}$ by the MC method, $\{\rho_{r}\}$
does not change strongly for various configurations of $\theta(r)$.
 
In the practical calculation, we employ the standard Metropolis algorithm 
with the local update\cite{Met}.
The typical sweep measurement is $(50 000-15 0000)\times$ (5 samples), 
and the acceptance ratio is 40-50$\%$. 
Errors are estimated from 20 samples by the jackknife method. 
Here we employ the {\em canonical ensemble}, and therefore 
the mean total number of boson, $\sum_r\rho_r$, is conserved during MC up-dates.
This situation is suitable for real experiments of the OL\cite{Bloch}.
We employ the periodic boundary condition and the lattice size is $L^{3}$, 
where $L$ is the linear system size.

%delta \tau
$\Delta\tau$ has a relation, $\Delta \tau L=1/(k_{B}T)$, 
thus its dimension is 1/(energy). In our calculation, we set $\Delta\tau=1$, 
regarded as unit of inverse of energy depending on system temperature.  
To identify the phase boundary, we calculate the internal energy $E$ and
``specific heat" $C$ that are defined as follows,
\begin{equation}
E=\langle H_{\rm qXY} \rangle/L^3, \;\;\;
C=\langle (H_{\rm qXY}-EL^3)^2\rangle/L^3,
\label{EC}
\end{equation}
where $\langle \cdots \rangle$ means the expectation value 
calculated as in Eq.(\ref{ZLqXY}),
\begin{eqnarray}
&&\langle \cdots \rangle \equiv  \int [d\rho_{r}]
\langle \cdots \rangle(\{\rho_{r}\}), \nonumber \\
&&\langle \cdots \rangle(\{\rho_{r}\})
= \int[d\theta (r)](\cdots)e^{-S^{\rm L}_{\rm qXY}}.
\label{expqXY}
\end{eqnarray} 

\begin{figure}[t]
\centering
\includegraphics[width=8cm]{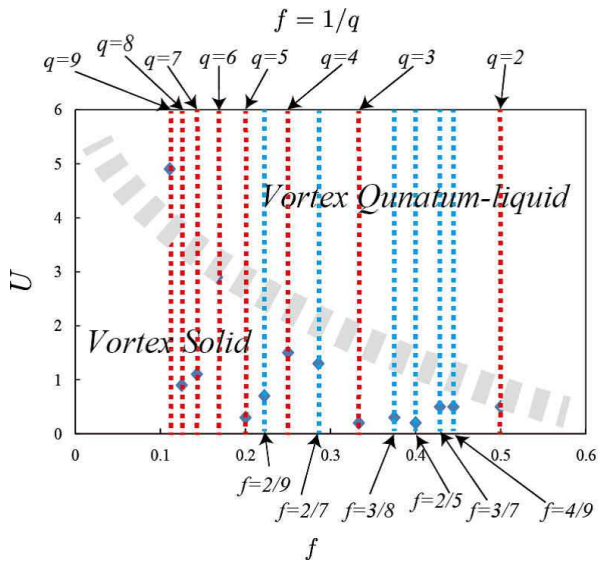}
\caption{(Color online) Phase diagram in $f$-$U$ plane.
The red-dotted lines indicates the $f=1/q$ vortex solid states
and the blue-dotted lines the $f=p/q$ vortex solid states.
The phase boundary between the solid states and the vortex quantum-liquid states
is determined by the behavior of vortex lines in the imaginary time direction.
See Fig.\ref{vortex_line}. $J=5$.
}
\label{PD}
\end{figure}

In Fig.\ref{PD}, we show the global phase diagram obtained by the eQMC for $J=5.0$,
and $\bar{\rho}=2.0$ ($\bar{\rho}$ is the mean density per site, i.e., 
$\bar{\rho}=\langle \rho_r \rangle$). 
The phase diagram is shown in the $(f-U)$ plane
and especially we focus on the magnetic flux regime $1/9 \leq  f \leq 1/2$.

%%%%%%%%%%%%%%%%%%%%%%%%%%%%%%%%%%
%Fig.1
%%%%%%%%%%%%%%%%%%%%%%%%%%%%%%%%%%%%%%%%%%%%%%%%%%%%%%%%%%%%%%
%%%%%%%%%%%%%%%%%%%%%%%%%%%%%%%%%
%Fig.2
\begin{figure}[t]
\centering
\includegraphics[width=7cm]{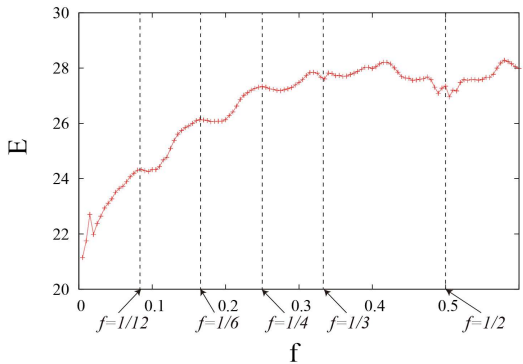}
\caption{(Color online) The internal energy $E$ as a function of the strength of
the magnetic flux $f$.
At $f\simeq 0.02$, there exists a sharp peak that indicates a phase 
transition.
Correlation function indicates that the system loses the SF at that point.
$J=5$ and $U=10$. 
Calculated $E$ indicates the existence of stable states (local minimums)
for certain values of $f$ like $f={1 \over 2}$ and  ${1 \over 3}$, whereas
at $f={1 \over 4}, {1 \over 6}$ and ${1 \over 12}$, a plateau appears.
System size $L=12$.
}
\label{XYE1}
\end{figure}
%%%%%%%%%%%%%%%%%%%%%%%%%%%%%%%%%%%%%%%%%%%%%%%%%%%%%%%%%%%%%%

By calculating the Bose correlation function $\langle a^\dagger_ra_{r'}\rangle$ 
by means of the eQMC, we have verified that for $f=0$ and a sufficiently 
large hopping $J$, a SF with the long-range order of the phase $\theta (r)$ forms, i.e.,
$\langle a^\dagger_ra_{r'}\rangle\rightarrow c\neq 0$ for $|r-r'|\rightarrow \infty$. 
As the value of $f$ is increased, however, the SF order disappears at $f\simeq 0.02$.
This implies that the critical magnetic field $B_c\sim 2\pi \times 0.02$.
The internal energy $E$ shown in Fig.\ref{XYE1} exhibits a sharp peak at $f\simeq 0.02$, 
and therefore the phase transition seems to be of first order. 
As the value of $f$ is increased furthermore, we found that at specific values
of $f$ like $f=p/q$, where $p$ and $q$ are co-prime integers, stable vortex-solid
states form. 
We searched such a state for $1/9 \leq  f \leq 1/2$ and identified fourteen states.
In the free electron lattice system in a uniform magnetic field, specific states for
general $f=p/q$ were predicted by Hofstadter\cite{Hofstadter}. 

\begin{figure}[t]
\centering
\includegraphics[width=8cm]{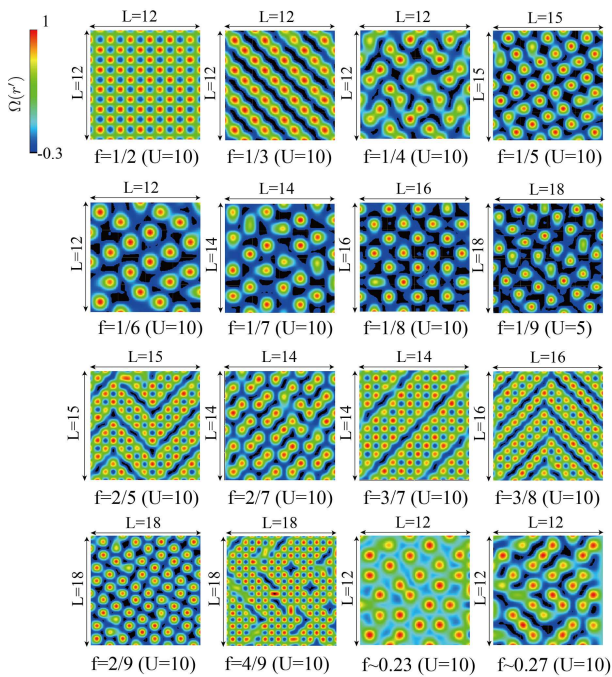}
\caption{(Color online) Snapshots of vortex $\Omega (r')$ for 
$\bar{\rho}=2.0$ and $J=5$. 
They show vortex solid states in $f=p/q$ ($1/9 \leq  f\leq 1/2$).
The results indicates 14 patterns of the vortex solid states, which are 
all of possible combinations of coprime numbers $p$ and $q$. 
For $f={1 \over 2}$ and ${1 \over 3}$, the rigid patterns appear, whereas
for $f={1 \over 4}$ and ${1 \over 6}$ locations of vortices are sightly loose.
On the other hand for incommensurate $f$'s (0.23 and 0.27), 
vortices do not form a lattice.
}
\label{vortex_solid}
\end{figure}

In Fig.\ref{PD}, the observed vortex solid states are indicated in the red- and 
blue-dotted lines for $J=5$ and their snapshots are shown in 
Fig.\ref{vortex_solid}.
We define vorticity $\Omega(r')$ on the dual site $r'$ of the site $r$
as follows:
\begin{eqnarray}
\Omega (r') &=& \frac{1}{4}\biggl[\sin(\theta(r+\hat{x})-\theta(r)+A_{\hat{x}}(r))\nonumber\\
&+&\sin(\theta(r+\hat{x}+\hat{y})-\theta(r+\hat{x})+A_{\hat{y}}(r+\hat{x}))\nonumber\\
&+&\sin(\theta(r+\hat{y})-\theta(r+\hat{x}+\hat{y})-A_{\hat{x}}(r+\hat{y}))\nonumber\\
&+&\sin(\theta(r)-\theta(r+\hat{y})-A_{\hat{y}}(r))\biggr].
\end{eqnarray} 
Expectation value of $\Omega(r')$ is calculated by the eQMC rather 
straightforwardly. 

\begin{figure}[h]
\centering
\includegraphics[width=8.5cm]{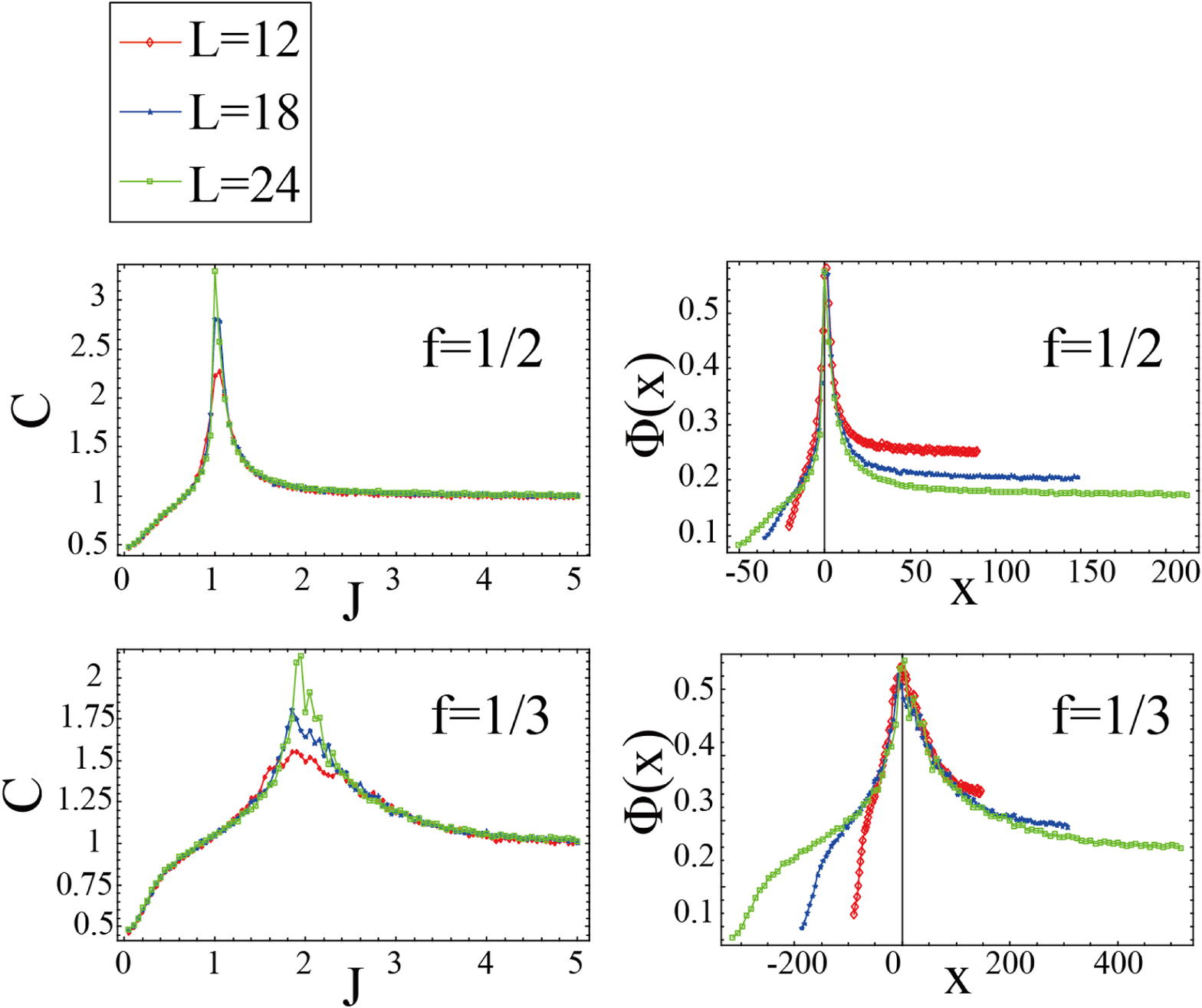}
\caption{(Color online) Specific heat $C$ for the system size $L=12,18$ and $24$.
Scaling function of the FSS for $f={1 \over 2}$ and 
$f={1 \over 3}$.}
\label{FSS1}
\end{figure}

We have found that the formation of vortex solid state depends on the 
lattice size $L$; 
i.e., for the $f=p/q$ vortex solid state to form, the spacial lattice size must be 
$qN \times qM$ ($N$ and $M$ are natural numbers).
As seen in Fig.\ref{vortex_solid}, the quantized vortices are pinned 
at sites of the dual lattice of the OL and they form solid pattern.
In our previous study\cite{KKK0} by the Monte-Carlo simulation assuming a 
homogeneous density of boson, some specific vortex solids (e.g., $f=1/2,1/3$ and $1/4$) 
were found.
However in the present study using the eQMC, we found more general 
solid patterns $f=p/q$.
This is due to the fact that the spatial density modulations are included as 
the variational parameter $\{\rho_{r}\}$ in the eQMC. 
As the density fluctuations suppress the fluctuations of the phase degrees of freedom
through the quantum uncertainty relation, and then the stable vortex solid 
formation is enhanced in the present study. 
Behavior of the ``specific heat" $C$ as a function of the hopping $J$ 
for $f={1 \over 2}$ and ${1 \over 3}$ is shown in Fig.\ref{FSS1}.
The results indicate the second-order phase transitions from the vortex solid 
to the disordered state as $J$ is decreased.
 
In Fig.\ref{vortex_solid}, we also show snapshots for incommensurate (`irrational')
magnetic flux $f\sim 0.23$ and $0.27$.
It is obvious that vortices are located in sites of the dual lattice but they do not
form a regular crystalline pattern.
However they are rather stable against the MC updates.
Therefore we conclude that an {\em amorphous state} of vortex forms at such fillings,
although at the fillings belonging to the plateaus in Fig.\ref{XYE1}, 
a regular vortex lattice is to be maintained by the lattice pining effect.

%%%%%%%%%%%%%%%%%%%%%%%%%%%%%%%%%
%Fig.3
%%%%%%%%%%%%%%%%%%%%%%%%%%%%%%%%%%%%%%%%%%%%%%%%%%%%%%% 
%%%%%%%%%%%%%%%%%%%%%%%%%%%%%%%%%
%Fig.4
%%%%%%%%%%%%%%%%%%%%%%%%%%%%%%%%%%%%%%%%%%%%%%%%%%%%%%%
%%%%%%%%%%%%%%%%%%%%%%%%%%%%%%%%%
%Fig.5
\begin{figure}[h]
\centering
\includegraphics[width=8cm]{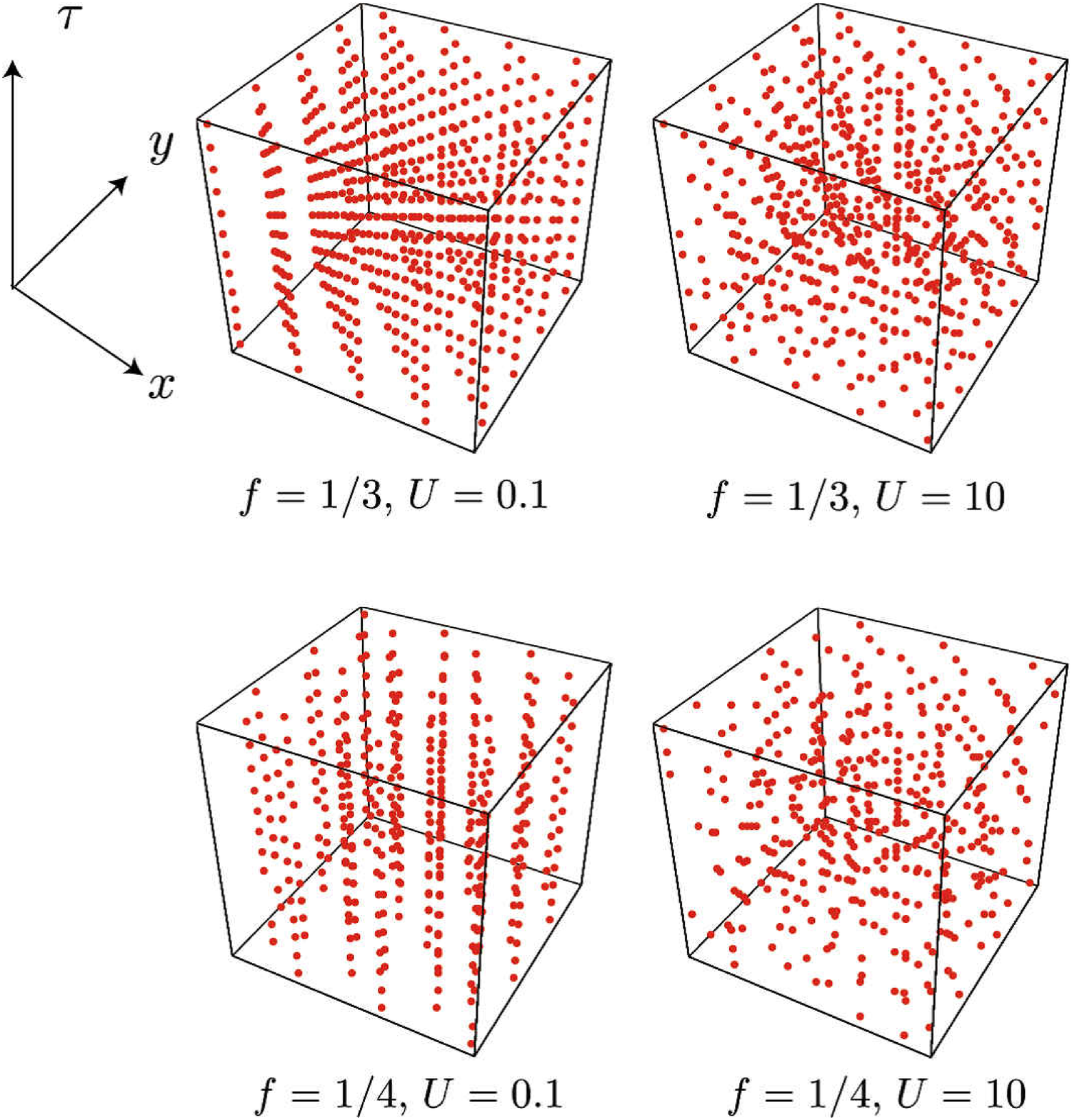}
\caption{(Color online) The typical vortex lines in $f=1/3$ and $1/4$ along the imaginary time direction $\tau$.
The red dots are the vortices residing in the dual site $r$.
As the on-site repulsive interaction $U$ increases,
the vortex-lines are entangled, and the quantum tunneling from the vortex lattice
shown in Fig.\ref{vortex_solid} to another parallel-translated one takes place.
When $U$ is small, the 
vortex solid states does not change along imaginary time direction.}
\label{vortex_line}
\end{figure}
%%%%%%%%%%%%%%%%%%%%%%%%%%%%%%%%%%%%%%%%%%%%%%%%%%%%%%%

As we explained above, the density fluctuation influences the dynamics of the
phase degrees of freedom $\theta(r)$.
Therefore it is interesting to see how the vortex behavior changes with
the strength of the on-site repulsion $U$.
Numerical simulation for the $f=p/q$ vortex solid state shows that  
as the value of $U$ increases, 
the vortex-lines in the imaginary-time direction starts to fluctuate. 
In Fig.\ref{vortex_line}, we show the typical vortex-lines in the $f=1/3$ and $1/4$ cases.
When $U$ is small, $U=0.1$, the vortex-lines are straight in the imaginary time direction,
whereas for large $U$, $U=10$, the vortex-lines are entangled with each other, 
i.e., the regular vortex lattices shown in Fig.\ref{vortex_solid} at some spatial
layer tend to distort in adjacent layers and then the parallel-translated
regular lattices reappear in some other layers.
This different behavior of the vortex line stems from the quantum fluctuation,
i.e., whether the degenerate vortex solid states in 2D are superposed or not
through {\em quantum tunneling} processes.
When the degenerate states in 2D are superposed, 
the state may be regard as a {\em vortex quantum-liquid} state.
As shown in the phase diagram in Fig.\ref{PD}, the phase boundary between
the vortex solid phase and the vortex quantum-liquid phase exists.
However the phase boundary is not clear because the $U$-dependent term 
in $S^{\rm L}_{\rm qXY}$ in Eq.(\ref{lattice_qXY}) generates only one-dimensional effect. 
Here we conclude that {\em the on-site interaction $U$ melts the rigid vortex-line states
into the fluctuating vortex line states}. 
In following section, the observed phenomenon in the above is studied
by using a duality transformation of the qXYM.  

Finally, by using the finite-size scaling (FSS) of $C$, we numerically
obtain the values of the critical exponents.
See Fig.\ref{FSS1}.
By the FSS hypothesis, $C$ for the system size $L$, $C_L(\epsilon)$, is
parametrized as
\begin{equation}
C_L(\epsilon)=L^{\sigma/\nu}\Phi(L^{1/\nu}\epsilon^\nu),
\label{FSS}
\end{equation}
where $\Phi(x)$ is a scaling function and $\epsilon=(J-J_\infty)/J_\infty$
with the critical coupling $J_\infty$ for the system $L\rightarrow \infty$.
In Eq.(\ref{FSS}), $\sigma$ and $\nu$ are critical exponent, in particular,
$\nu$ is the critical exponent of the correlation length 
$\xi \propto |J-J_\infty|^{-\nu}$.
The results of the FSS are shown in Fig.\ref{FSS1} for $f=1/2$ and $1/3$.
The critical exponents are estimated as
$\nu=0.80 (0.55), \sigma=0.44 (0.24)$ for $f=1/2 (1/3)$

%%%%%%%%%%%%%%%%%%%%%%%%%%%%%%%%%%%%%%%%%%%%%%%%%%%%%%%

\section{Duality transformation}

In this section, we apply a duality transformation for the qXYM to understand 
the behavior of vortex observed in the previous section by the eQMC.
This approach is an extension of the famous analysis on the 2D classical XY 
model, see for example Ref.\cite{Savit,Nagaosa}.
Duality-transformed model of the qXYM is described in term of the vortex density 
and topological current field.
Besides the vortex-solid formation, it explains the vortex-line fluctuations in the
$\tau$-direction controlled by the on-site repulsion $U$.
Furthermore from the dual model, we find the $f=1/q$ rule for vortex solid pattern,
which explains how the vortex-solid pattern is determined by the magnetic field $f$, 
the lattice size $L$, and vortex density-density interaction.

%%%%%%%%%%%%%%%%%%%%%%%%%%%%%%%%%%%%%%%%%%%%%%%%%%%%%%%%%%%%
\subsection{Derivation of the dual-qXYM}

To derive the dual-qXYM, let us focus on the first and
second terms in Eq. (\ref{lattice_qXY}), and define
\begin{eqnarray}
Z_{\tau +hop}&=&\int [d\theta(r)] \ e^{-S_{\tau +hop}},
\label{Z_tau}
\end{eqnarray}
%with
\begin{eqnarray}
S_{\tau +hop}&=&-\sum_{r,\tau}\frac{1}{4U}\cos (\theta(r+\tau)-\theta(r) )\nonumber\\
&&-\sum_{r,\mu}J\cos (\theta(r+\mu)-\theta(r)+A_{\mu}(r)),
\label{Stauhop}
\end{eqnarray} 
where we have set $\Delta\tau\equiv 1$, $\rho_{r}= \langle \rho_{r} \rangle = 1$ 
for simplicity.
(Effects of fluctuations in the local density $\rho_r$ will be discussed in Sec.IV.C and 
Sec.IV.D.)
In the following, we express the action $S_{\tau +hop}$ as follows for notational simplicity,
\begin{eqnarray}
S_{\tau +hop} = -\sum_{r,\mu}J_{\mu}\cos (\theta(r+\mu)-\theta(r)+A_{\mu}(r)),
\end{eqnarray} 
where we have redefined direction labels, the couplings, and vector potentials as 
\begin{equation}
\mu=0,1(\hat{x}), 2(\hat{y}),  \; J_{0}=\frac{1}{8U}, \; J_{1}=J_2=J
\label{3DXY1}
\end{equation}
and use the gauge
\begin{equation}
A_{0}(r)=0, \; A_{1}(r)=2\pi f y, \; A_{2}(r)=0.
\label{3DXY2}
\end{equation} 
\vspace{0.3cm} 

Let us apply the periodic Gaussian approximation to Eq.(\ref{Z_tau})\cite{Savit,Nagaosa},
\begin{eqnarray}
&&e^{J_{\mu}\cos(\theta(r+\mu)-\theta(r)+A_{\mu}(r))} \nonumber \\
&&\rightarrow \sum^{\infty }_{n_\mu=-\infty }
e^{J_{\mu}}\exp\biggl[-\frac{J_{\mu}}{2}\Big(\theta(r+\mu)-\theta(r)  \nonumber   \\
&&\hspace{2cm} +A_{\mu}(r)-2\pi n_{\mu}(r')\Big)^{2}\biggr],
\label{periodicity1}
\end{eqnarray} 
where the vector field $n_{\mu}(r')$ is defined on the sites $r'$ of the 3D dual lattice,
and this field represents $2\pi$ periodicity of the cosine term in $S_{\tau +hop}$. 
Substituting Eq. (\ref{periodicity1}) into the partition function Eq.(\ref{Z_tau}), 
we obtain the following equation by using Poisson formula 
(see for example, Ref.\cite{Savit}),
\begin{widetext}
\begin{eqnarray}
Z_{\tau +hop}&\propto&\int[d\theta(r)]\sum_{l_{\mu}(r')}\exp\biggl[ -\sum_{r,\mu}\biggl(\frac{l^{2}_{\mu}(r')}{2J_{\mu}}-il_{\mu}(r')(\theta(r+\mu)-\theta(r)+A_{\mu}(r))\biggr)\biggr]\nonumber\\
&\propto&\sum_{l_{\mu}(r')}\exp \biggl(-\sum_{r,\mu}\frac{l^{2}_{\mu}(r')}{2J_{\mu}}-il_{\mu}(r)A_{\mu}(r)\biggr)\Pi _{r}\delta(\sum_{\mu}l_{\mu}(r')-l_{\mu}(r'-\mu)),
\label{HST}
\end{eqnarray}
\end{widetext}
where we have used the Hubbard-Stratonovich transformation, and 
the vector integer field $l_{\mu}(r')$ has been introduced as the boson-current variables.
To solve the $\delta$-function constraint in Eq.(\ref{HST}), we introduce new 
integer-gauge fields $\tilde{a}_{\mu}(r)$, with which $l_{\mu}(r')$ is expressed as
$l_{\mu}(r')\equiv \frac{1}{2\pi}
\epsilon_{\mu\nu\lambda }\nabla_{\nu}\tilde{a}_{\lambda }(r')$.
By substituting the above expression of $l_{\mu}(r')$ into the action,
we have
\begin{widetext}
\begin{eqnarray}
Z_{\tau +hop}&\propto&\sum_{\tilde{a}_{\mu}(r)}\exp \biggl(-\sum_{r,\mu}\frac{(\epsilon _{\mu\nu\lambda}\nabla _{\nu} \tilde{a}_{\lambda }(r))^{2}}{4\pi J_{\mu}}
-i\epsilon _{\mu\nu\lambda}\nabla _{\nu} \tilde{a}_{\lambda }(r)A_{\mu}(r)\biggr),
\label{new_gauge}
\end{eqnarray}
\end{widetext}
where the operator $\nabla_{\mu}$ is lattice nabla, defined by 
$\nabla_{\mu}A_{\nu}(r)\equiv A_{\nu}(r+\mu)-A_{\nu}(r)$.
Here we use again Poisson summation formula, and then the integer variables
$\tilde{a}_{\mu}(r')$ are transformed into the continuum variables $\phi_{\mu}(r')$,
and the partition function is expressed as,
\begin{widetext}
\begin{eqnarray}
Z_{\tau +hop}&\propto&\int^{\infty }_{-\infty }[d\phi_{\mu}(r')]\sum_{m_{\mu}(r')}
\exp \biggl(-\sum_{r,\mu}\frac{(\epsilon _{\mu\nu\lambda}
\nabla _{\nu} \phi_{\lambda }(r'))^{2}}{4\pi J_{\mu}}
-i\epsilon _{\mu\nu\lambda}\nabla _{\nu} \phi_{\lambda }(r')A_{\mu}(r)
+2\pi im_{\mu}(r')\phi_{\mu}(r')\biggr)\nonumber\\
&=&\int^{\infty }_{-\infty }[d\phi_{\mu}(r')]\sum_{m_{\mu}(r')}
\exp \biggl(-\sum_{r,\mu}\frac{(\epsilon _{\mu\nu\lambda}
\nabla _{\nu} \phi_{\lambda }(r'))^{2}}{4\pi J_{\mu}}
-2\pi if_{\mu} \phi_{\mu}(r')+2\pi im_{\mu}(r')\phi_{\mu}(r')\biggr)  \nonumber  \\
&\equiv& \sum_{m_{\mu}(r')}\int^{\infty }_{-\infty }[d\phi_{\mu}(r')]
 e^{-S_{v}},  
\label{partitionSv}
\end{eqnarray}
with
\begin{eqnarray}
S_{v}(\tilde{j}_{\mu}(r'),\phi_{\mu}(r'))=\sum_{r',\mu}\biggl[-\frac{1}{4\pi J_{\mu}}
(\epsilon _{\mu\nu\lambda }\nabla _{\nu}\phi_{\lambda }(r'))^{2}
+i\phi_{\mu }(r')\tilde{j}_{\mu}(r')\biggr],
\label{Sv1}
\end{eqnarray}
\end{widetext} 
where we have introduced the three-component flux field $f_\mu(r)$
for the notational simplicity
$2\pi f_{0}(r)=\epsilon _{0ij}\nabla _{i}A_{j}(r), \ f_{1(2)}(r)=0$, and 
$\tilde{j}_{\mu}(r')=2\pi(m_\mu(r')-f_\mu(r))$.  
By integrating the ``gauge fields" $\phi_{\mu}(r')$, we obtain the final expression of 
the dual model. 

As the action $S_v$ in Eq.(\ref{Sv1}) is invariant under a local gauge transformation
$\phi_\lambda(r')\rightarrow \phi_\lambda(r')-\nabla_\lambda\alpha(r')$ with
an arbitrary scalar function $\alpha(r')$, we have to fix the gauge of
$\phi(r')$ to integrate out $\phi(r')$.
To this end, we consider the case of $J_\mu=J$ for simplicity and 
employ the Lorentz gauge $\sum_{\mu=0,1,2}\nabla_\mu\phi_\mu=0$.
In the present path-integral formalism, this gauge condition is easily 
imposed by adding the gauge-fixing term $(\sum_{\mu}\nabla_\mu\phi_\mu)^2$
to the action $S_v$.
In the Lorentz gauge,
\begin{eqnarray}
&&S_v-{1 \over 4\pi J}(\sum_{r',\mu}\nabla_\mu\phi_\mu)^2
+(\tilde{j}-\mbox{terms})  \nonumber \\
&& \hspace{1cm}
={1 \over 4\pi J}\sum_{r',\mu,\lambda}\phi_\lambda\nabla^2_\mu\phi_\lambda 
+(\tilde{j}-\mbox{terms}).
\label{Sv2}
\end{eqnarray}
Integration over $\phi(r)$ in Eq.(\ref{partitionSv}) can be performed straightforwardly
and obtain the dual model of the vortex density and vortex current $\tilde{j}_\mu(r')$
as follows,
\begin{eqnarray}
&&\sum_{m_{\mu}(r')}\int^{\infty }_{-\infty }[d\phi_{\mu}(r')]e^{-S_{v}}
=\sum_{m_{\mu}(r')}e^{-S_{\rm dual}},  \nonumber \\
&&S_{\rm dual}=-{\pi \over J}\sum_{r',r''}\tilde{j}_\mu(r)
(\sum_\mu\nabla^2_\mu)^{-1}(r,r')
\tilde{j}_\mu(r').
\label{Sdual1}
\end{eqnarray}

In what follows, we shall study {\em the stationary state} of the vortex configurations
like the vortex solid.
We first interested in the vortex density-density interaction in the 
dual model, i.e., the interaction between the component $\tilde{j}_{0}$ 
in Eq.(\ref{Sdual1}), and evaluate the corresponding term as
\begin{eqnarray}
-\pi J\sum_{r'}\tilde{j}_{0}(r)\frac{1}{\nabla^{2}_{1}+\nabla^{2}_{2}}\tilde{j}_{0}(r')
=\pi J\sum_{\rm B.Z.}\tilde{j}_{0}(k)\frac{1}{|\tilde{k}|^{2}}\tilde{j}_{0}(-k), \nonumber\\
\label{jjterm}
\end{eqnarray}
where 
$\tilde{j}_{\mu}(r)=\sum_{\rm B.Z.}\tilde{j}_{\mu}(k)e^{ikr}$ 
(B.Z. refers to the first Brillouin Zone) and
$\tilde{k}_{\mu}=\frac{2}{a_L}\sin\frac{k_{\mu} a_{L}}{2}$.
The same result with Eq.(\ref{jjterm}) is obtained in the Coulomb gauge
$\sum_{\alpha=1,2}\nabla_\alpha\phi_\alpha=0$.
In the long-distance region $|r'-r''| \gg a_L$, Eq.(\ref{jjterm}) behaves as
\begin{eqnarray}
&&\pi J\sum_{\rm B.Z.}\tilde{j}_{0}(k)\frac{1}{|\tilde{k}|^{2}}\tilde{j}_{0}(-k)\nonumber\\
&\sim&\pi J\sum_{r',r''}\tilde{j}_{0}(r')\sum_{k}\frac{1}{k^{2}}e^{-ik(r'-r'')}\tilde{j}_{0}(r'')\nonumber\\
&\propto& \pi J\sum_{r',r''}\tilde{j}_{0}(r')\log|r'-r''|\tilde{j}_{0}(r'')
+\alpha\biggl(\sum_{r'}\tilde{j}_{0}(r')\biggr)^{2}\nonumber\\
&=& \pi J\sum_{r',r''}(j_{0}(r')-2\pi f)\log|r'-r''|(j_{0}(r')-2\pi f)  \nonumber\\
&+&\alpha\biggl(\sum_{r'}(j_{0}(r')-2\pi f)\biggr)^{2}.
\label{j0j0}
\end{eqnarray}
%\end{widetext}
where $j_{\mu}(r')\equiv 2\pi m_\mu(r')$, and
we have introduced the $\alpha$-term by regularizing the infrared divergence
in the $k$-integration and then $\alpha\sim \ln(L/a_L)$. 
It is easily verified that the current-current correlations $({j}_1-{j}_1)$ and
$({j}_2-{j}_2)$-terms have a similar form with Eq.(\ref{j0j0}), but
its coefficient is $1/4U$.
Therefore as $U$ is getting large, the current correlation becomes weak
and nontrivial configurations of the current $\vec{j}=(j_1,j_2)$ appear.
This means the correlation of vortex density in the $\tau$-direction is getting weak
as we observed in the numerical simulations in Fig.\ref{vortex_line}.
In following section, we explain the generated patterns of the vortex solid 
by the eQMC from Eq.(\ref{j0j0}).

%%%%%%%%%%%%%%%%%%%%%%%%%%%%%%%%%%%%%%%%%%%%%%%%%%%%%%%%%%%%%%%%

\subsection{Vortex solid $f=1/q$ rule} %\label{vortexsolid}

In previous section, we derived the dual-qXYM model.
For the stationary configurations, the vortex-density interaction 
has the logarithmic form similar to that of 2D XY model, and its coefficient
is proportional to the hopping amplitude $J$. 
Moreover, the infrared singularity gives the ``charge-neutrality condition" as
\begin{equation}
(\sum_{r'}\tilde{j}_0(r'))^2 \propto
(\sum_{r}(2\pi m_{0}(r)-2\pi f))^{2}\rightarrow 0.
\label{CNC}
\end{equation}
Thus for sufficiently large hopping $J$, the BEC is realized in the system and
the above condition decides the total number of vortex in the system
(more precisely, the number of (vortex $-$ anti-vortex)). 
Here, we again show the vortex-density interactions,
\begin{eqnarray}
&&J\sum_{r,r'}(2\pi m_{0}(r)-2\pi f)\log|r-r'|(2\pi m_{0}(r')-2\pi f)\nonumber\\
&+&\alpha(\sum_{r}(2\pi m_{0}(r)-2\pi f))^{2}.
\label{vortex_interaction}
\end{eqnarray} 
From the above interactions, we qualitatively explain the patterns of vortex-solid
observed by eQMC in the previous section.
For simplicity, we shall consider the case of the magnetic flux $f=1/q$.
\begin{enumerate}
\item In the neutrality condition Eq.(\ref{CNC}), $f=1/q$ and 
$m_0(r)/2\pi$ is an integer.
$q$ adjacent plaquettes have to be grouped for making $q\times 1/q=1$
flux quantum.
To achieve this globally, the lattice system size has to be $qN_{x}\times qN_{y}$ 
($N_{x}$ and $N_{y}$ are natural numbers). 
This is the lattice size condition of the system.
\item The lattice system is divided into a number of $q$-adjacent plaquettes.
One quantized vortex resides in arbitrary one of $q$-adjacent plaquettes.
To realize low-energy configurations,
the vortex density interactions have to be minimized. 
This decides the distribution pattern of a number of vortices.
\item The number of the distribution pattern corresponds 
to the number of the degenerate state of the $f=1/q$ vortex-solid state.
When the particle density is high enough, a superposed vortex-solid
patterns is realized because of the long-range repulsion between vortices.
\end{enumerate}

The above consideration can be easily extended to the more general cases with
$f=p/q$ ($p$ and $q$ are co-prime), and from this consideration 
it is expected that all possible vortex solids can be observed by the present 
eQMC simulations

%%%%%%%%%%%%%%%%%%%%%%%%%%%%%%%%%%%%%%%%%%%%%%%%%%%%%%%%%%%%%
\subsection{Phase diagram in dilute boson regime}

\begin{figure}[t]
\centering
\includegraphics[width=8cm]{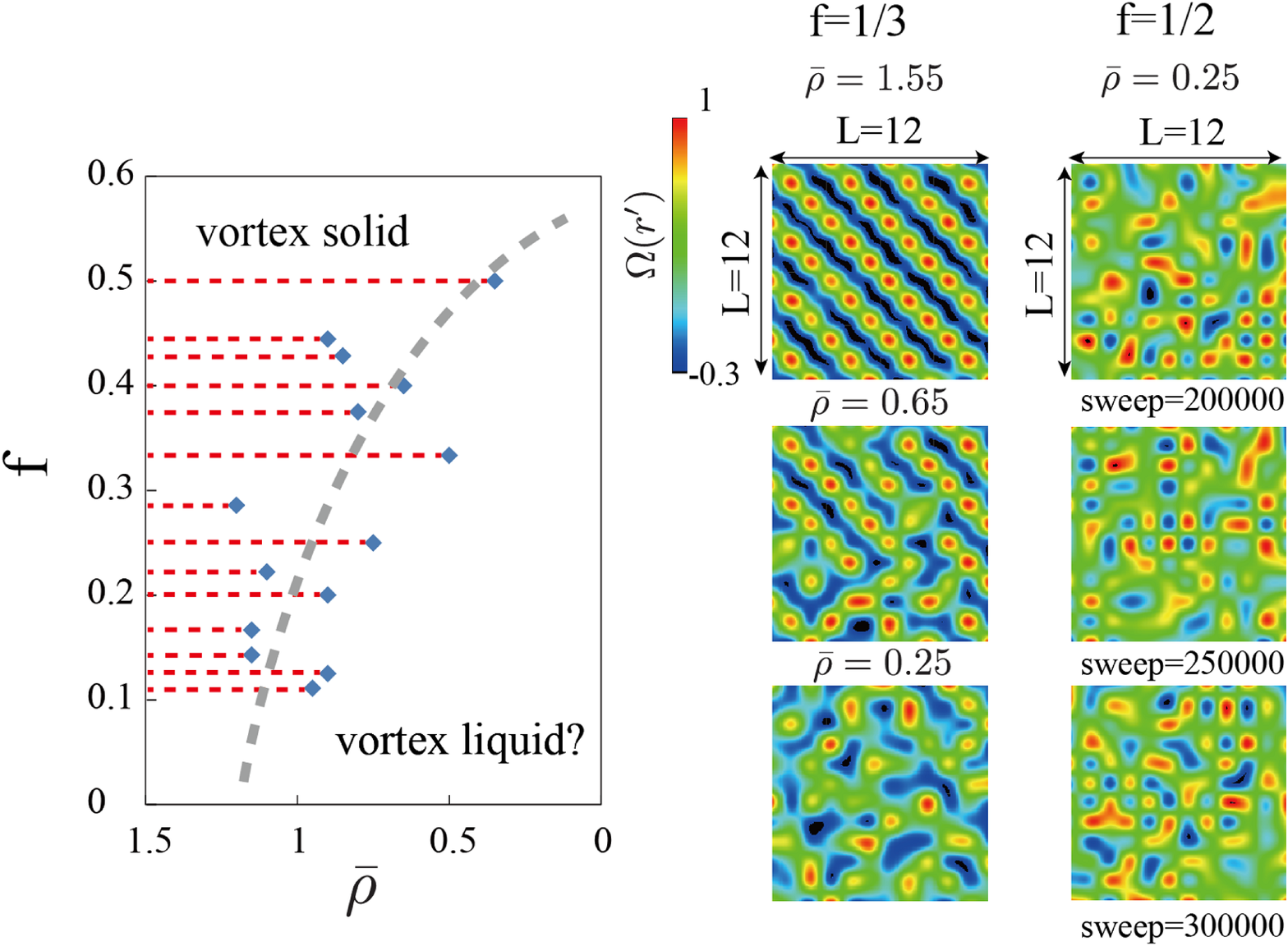}
\caption{Numerical results of the dilute density regime. (Left panel)
The phase diagram in the $\bar{\rho}-f$ plane.
The red broken lines show the observed vortex solid states in which 
the vortices are pinned on the dual lattice. 
(Right panel) Snapshots for $f=1/3$.
At $\bar{\rho}=2$, the robust vortex solid forms.
As the density is decreased, `melting' of solid takes place. 
At present, we do not have clear understanding of the states in the dilute regime.
However, we expect that they are a kind of vortex-liquid state. 
In the right column, we show the results for $f=1/2$ and $\bar{\rho}=0.25$, in
which a bosonic fractional Hall state is expected to form.
The state is rather homogeneous but slightly unstable against the MC updates.
}
\label{dilute_PD}
\end{figure}

In Fig.\ref{dilute_PD}, we show the phase diagram of  vortex state in the 
low density regime.
As explained in Ref.\cite{dilute}, recent experiments can decrease the atomic
density without losing the phase coherence by applying a micro-wave to local
regimes of the OL and blowing away excited atoms from the OL.
By eQMC calculation, we found that as the average density of bosons is decreased, 
the vortex-solid states disappear.
As shown in Fig.\ref{dilute_PD}, locations of vortices do not exhibit any 
spatial patterns in the low-density region and the transition may be interpreted as
a melt of the vortex solid to a vortex liquid\cite{Cooper}.
Specific heat $C$ shows a sharp peak at the transition point for some values of $f$,
but for other values of $f$ there is only a moderate peak in $C$. 
Due to the diluteness of bosons,  the effect of the lattice is weekend 
and the system approaches to the continuum system.
Pining of vortex by the lattice is less effective, and the topological vortex number
is spread rather wide spatial regions.
These findings are in agreement with the intuitive picture such that 
for the low-density limit, the lattice works simply as a mesh employed for numerical study. 
In fact, the low-density limit means small hopping parameter 
$J(\rho_r\rho_{r+\mu})^{1/2}$ and its relatively large local fluctuations, and then
the interaction term in Eq.(\ref{vortex_interaction}) becomes less effective.
This observation is quite instructive for the discussion on the realization
of the bosonic Laughlin state in the following section.

%%%%%%%%%%%%%%%%%%%%%%%%%%%%%%%%%%%%%%%%%%%%%%%%%%%%%%%%%%%%%%%%%%
%Fig.6

%%%%%%%%%%%%%%%%%%%%%%%%%%%%%%%%%%%%%%%%%%%%%%%%%%%%%%%%%%%%%%%%%%%

\subsection{Bosonic Laughlin state and Chern-Simons theory of the
Bose Hubbard Model}

In recent years, it was suggested that a state similar to the bosonic Laughlin
state forms in the Bose gas system loaded on the OL.
In the first-quantization language and also in the continuum space, 
the bosonic Laughlin state is given by the following wave function\cite{Laughlin},
\begin{equation}
\Psi(z_1,z_2, \cdots, z_N)=\prod_{j<k}(z_j-z_k)^m e^{-{1 \over 4}\sum|z_j|^2},
\label{BLS}
\end{equation}
where $z_j=x_j+iy_j$ and $m$ is an even positive integer.
From Eq.(\ref{BLS}), it is obvious that bosons in the Laughlin state 
have the hard-core nature as $\Psi \rightarrow 0$ for $z_j\rightarrow z_k \; (j\neq k)$.
In the previous work\cite{Sorensen,Hafezi}, a numerical diagonalization of a small system 
showed that
the ground-state of the hard-core BHM has a good overlap with the wave function
in Eq.(\ref{BLS}) for a weak magnetic field $f<0.2$ and low density $\bar{\rho}<0.1$
case.

It is interesting to ask if a state that has similar properties to 
the Laughlin state forms in the boson gas system on the OL for relatively large $f$'s
and particle density.
Such a state has a certain hidden order and low-energy excitations
behave like anyons.
To study this problem, Chern-Simons (CS) theory for the fractional quantum Hall 
effect (FQHE) is quite useful.
We previously introduced a lattice version of the CS theory that is well suited for
studying the present system\cite{IchinoseMatsui}.

By using this formalism, we transform the original boson operator $a_r$ in 
Eq.(\ref{BH}) to another boson operator $b_r$, which we call CS boson, 
by attaching $m$-magnetic flux to $a_r$,
\begin{eqnarray}
&&a_r=V_rb_r, \nonumber \\
&&V_r=\exp\Big[im\sum_{x'}\theta(r,x')(a^\dagger_{r'}a_{r'})\Big],
\label{CSboson}
\end{eqnarray}
where $x \ (x')$ denotes a site of the dual lattice paired to site $r \ (r')$ of the original 
lattice, and $\theta(r,x')$ is the azimuthal angle function on the lattice.
Please notice $a^\dagger_ra_r=b^\dagger_rb_r$.
Then, let us consider the specific case $f=m\bar{\rho}$ where $\bar{\rho}$
is again the average particle density.
In this case in terms of the CS boson $b_r$, the BHM Hamiltonian in Eq.(\ref{BH})
is expressed as,
\begin{eqnarray}
H_{\rm BH}=&&-J\sum_{r,\mu}b^\dagger_rW^\dagger_rW_{r+\mu}b_{r+\mu}
+\mbox{h.c.}  \nonumber   \\
&&+\sum_rU(b^\dagger_rb_r)^2,  \label{BHCS1} \\
W_r=&&\exp\Big[im\sum_{x'}\theta(r,x')(b^\dagger_{r'}b_{r'}-\bar{\rho})\Big],
\label{BHCS2}
\end{eqnarray}
where we have employed the symmetric gauge for $A_\mu(r)$, and used
the identity
$2\pi\epsilon_{\mu\nu}\nabla_\nu G(x,x')=\nabla_\mu\theta(r,x')$
[$\epsilon_{12}=-\epsilon_{21}=1, \epsilon_{11}=\epsilon_{22}=0$]
with the two-dimensional lattice Green function $G(x,x')$.
Beautiful CS gauge theory can be constructed for the system 
Eqs.(\ref{BHCS1}) and (\ref{BHCS2}) in the Lagrangian formalism,
but here we only discuss the possible mean-field (MF) solution to the ground-state
of the above system.

The MF analysis indicates the naive candidate for the ground-state of
$H_{\rm BH}$ in Eq.(\ref{BHCS1}) such as $b^{\rm MF}_r=\bar{\rho}^{1/2}$,
i.e., in which the BEC of the CS boson $b_r$ takes place.
From Eq.(\ref{CSboson}) and $f=m\bar{\rho}$, this MF solution accompanies
the homogeneous `condensation' of the $2\pi f$ flux quanta 
per plaquette to cancel out the external magnetic field.
This MF state corresponds to the bosonic Laughlin state and therefore in the
bosonic Laughlin state, the grand-state has the homogeneous particle
density $\bar{\rho}$ and also the homogeneous vortex density $2\pi f$
per site and per plaquette, respectively.
However for sufficiently large $\bar{\rho}$ and $f$, the above condition 
is not satisfied by the configuration of $a_r$ because the duality transformation
shows that vortex density at each plaquette $m_0(r')$ takes an integer value.
This is nothing but the lattice pining effect of the vortex.

On the other hand for small $\bar{\rho}$ and $f$, the fluctuations of the
parameter $J(\rho_r\rho_{r+\mu})^{1/2}$ cannot be neglected and
therefore the direct application of the result of the duality-transformation
is questioned.
It is quite useful to study, for example, the case with $f=0.5, \ \bar{\rho}=0.25$ and 
therefore $m=2$.
Snapshot of the topological number (vortex density) is shown in Fig.\ref{dilute_PD}.
It is obvious that a genuine homogeneous configuration is not realized there but
topological number tends to smear compared with the higher-density case
in Figs.\ref{vortex_solid} and \ref{dilute_PD}.
This result seems to  indicate the possibility of the bosonic Laughlin state
for very low particle density as indicated by the work on the very small systems
in Ref\cite{Sorensen,Hafezi}.
Methods in the present paper are also applicable for the study of such 
low density cases, i.e., the hard-core BH model.
To this end, the slave-particle representation for the hard-core boson 
is quite useful\cite{slave}.
This problem is under study  and we hope that the results will be 
reported in another publication in future.

%%%%%%%%%%%%%%%%%%%%%%%%%%%%%%%%%%%%%%%%%%%%%%%%%%%%%%%%%%%%%%%%%%%
\section{Haldane Bose-Hubbard model studied by $\mbox{e}$QMC}

Resent experiments on clod atoms succeeded in realizing 
the Haldane-Bose-Hubbard model 
in a honeycomb optical lattice\cite{Honeycomb_ex}.
This system is a bosonic analog of the celebrated Haldane model\cite{Haldane},
which is a fermionic system and possesses a topological phase due complex 
hopping amplitudes on the honeycomb lattice.
Besides the NN and NNN hopping terms, the on-site repulsion exists in the HBHM,
and as result of the competition of these three terms, the model has a rich
phase diagram. 
The HBHM realized by cold atomic gases on the optical lattice is highly controllable, e.g.,
besides the average particle density, the on-site repulsion, the hopping amplitude
and the artificial magnetic flux generated by the NN hopping can be controlled.

In this section, we shall study the HBHM by means of the eQMC and obtain 
the global phase diagram.
To this end, we calculate the internal energy, specific heat and certain
correlation functions to identify order of phase transitions and 
physical properties of phases.
By using the FSS, we obtain the critical exponents for phase transitions.
The previous work\cite{Hofstetter} clarified the phase diagram of the HBHM, 
however the system size in the numerical calculation (the exact diagonalization) 
is small, and only the case of unit filling was considered.
Therefore the present study using the eQMC is complementary to the previous
work.
In particular, besides those found in Ref.\cite{Hofstetter} we have found 
another phase boundary.

As in Ref.\cite{Hofstetter}, the Hamiltonian of the HBHM  is given by
\begin{eqnarray}
H_{\rm HBH}&=&-J_{1}\sum_{\langle i,j\rangle}a^{\dagger}_{i}a_{j}
-J_{2}\sum_{\langle\langle i,j\rangle\rangle}e^{-i\phi}a^{\dagger}_{i}a_{j}\nonumber\\
 &&+U\sum_{i}a^{\dagger}_{i}a_{i}a^{\dagger}_{i}a_{i},
\label{HBHM}
\end{eqnarray}
where $\langle i,j\rangle$ denotes NN sites, $\langle\langle i,j\rangle\rangle$ 
NNN sites of the honeycomb lattice.
The parameters $J_{1}$ and $J_{2}$ are the NN and NNN hopping amplitudes, 
respectively, and $\phi$ is a constant phase of the NNN hopping. 
By a calculation similar to that in Sec.II, the effective action for the HBHM is 
obtained as
\begin{eqnarray}
S_{\rm HBH}&=&\int d\tau \biggr(- \sum_{i}\frac{1}{4U}(i\partial_{\tau}\theta (i))^{2}\nonumber\\
&-&J_{1}\sum_{\langle i,j\rangle}\sqrt{\rho_{i}\rho_{j}}
\cos(\theta_{i}-\theta_{j})\nonumber\\
&-&J_{2}\sum_{\langle\langle i,j\rangle\rangle}
\sqrt{\rho_{i}\rho_{j}}\cos(\theta_{i}-\theta_{j}+\phi) \biggl).
\label{eff_HBHM}
\end{eqnarray}    

%%%%%%%%%%%%%%%%%%%%%%%%%%%%%%%%%
%Fig.7
\begin{figure}[h]
\centering
\includegraphics[width=5cm]{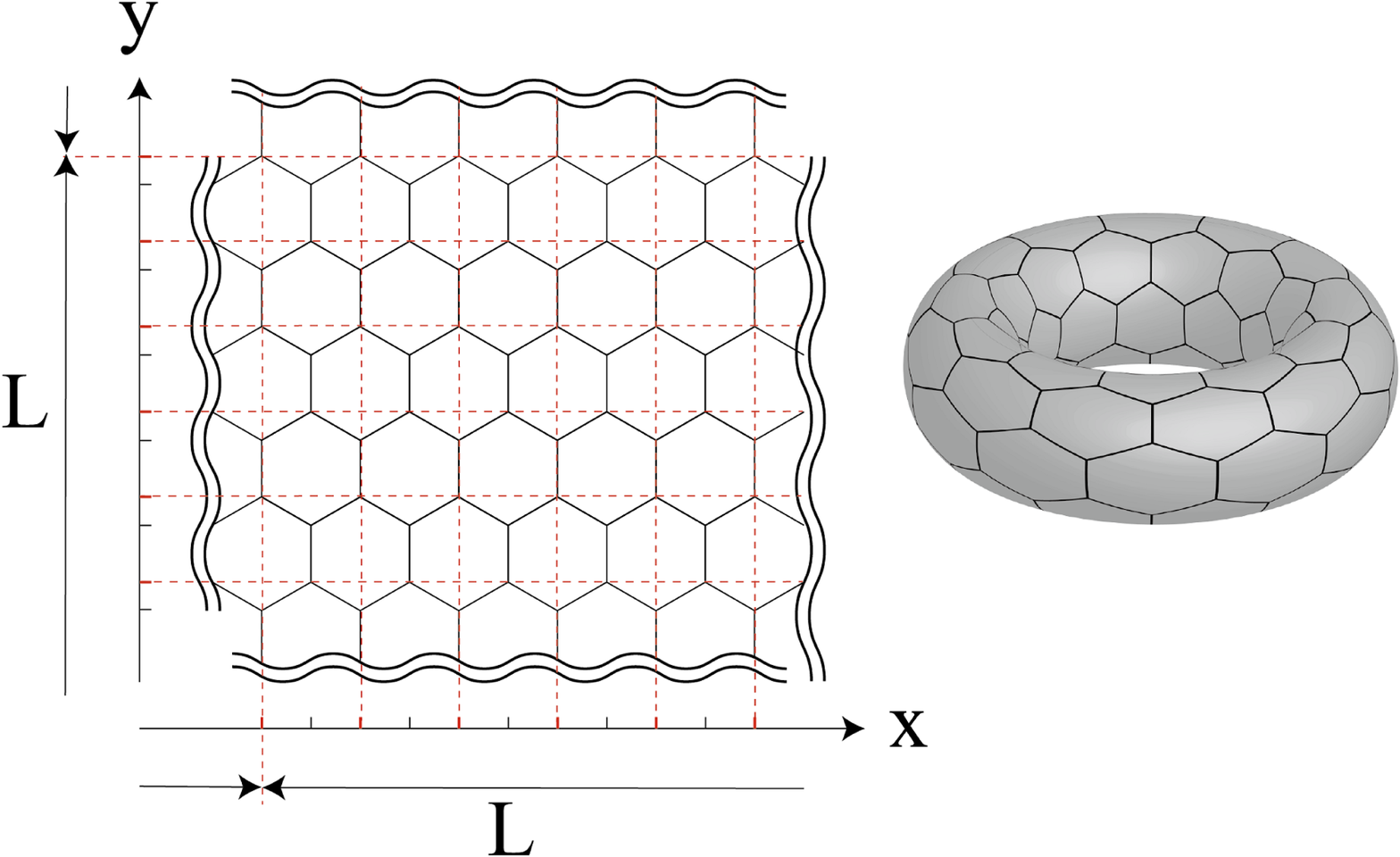}  
\includegraphics[width=5cm]{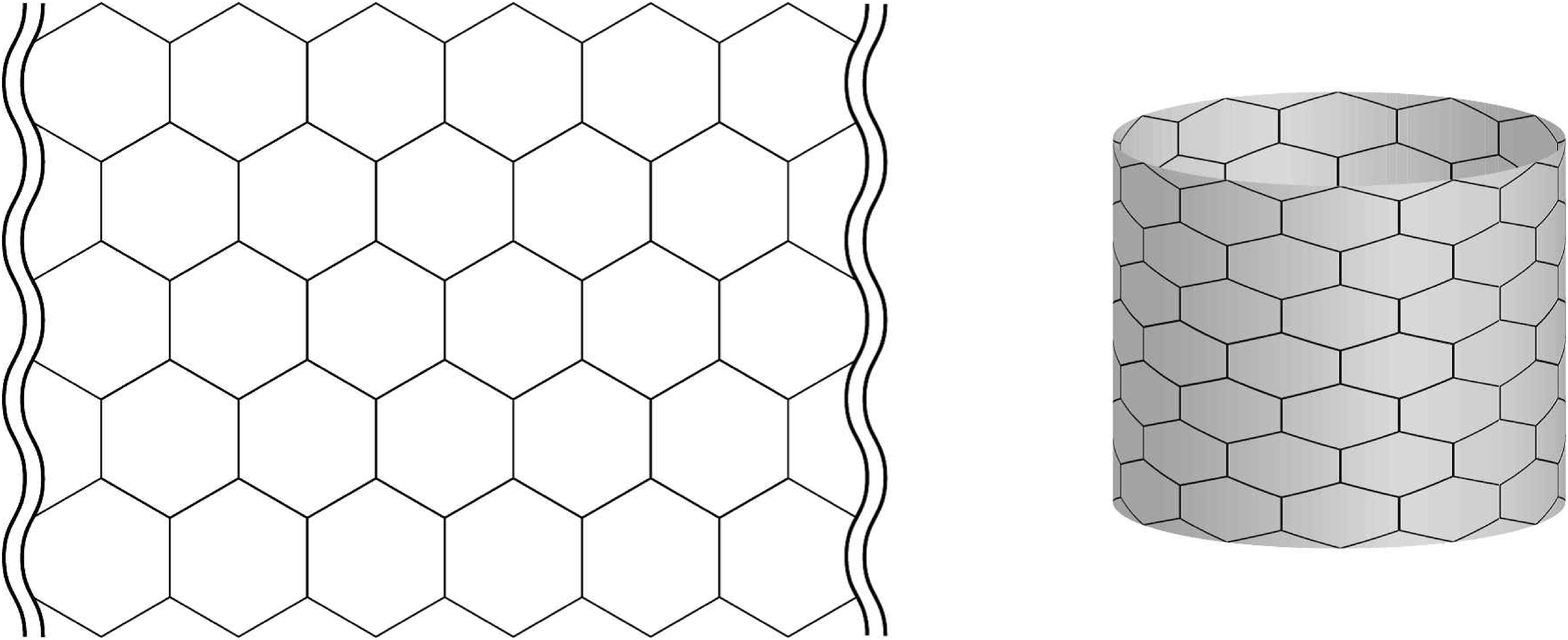}
\caption{(Color online) Honeycomb lattice with the periodic boundary
condition (torus) and also that of the cylinder geometry.
%%%%
The system size can be defined from A sub-lattice size, its lattice is a square $L\times L$ lattice.
The imaginary-time $\tau$ lattice is fixed $L_{\tau}=8$ and has periodic boundary condition. 
}
\label{honeycomb1}
\end{figure}
%%%%%%%%%%%%%%%%%%%%%%%%%%%%%%%%%%%%%%%%%%%%%%%%%%%%%%%%%%%%%%%%%%%%
%Fig.8
\begin{figure}[h]
\centering
\includegraphics[width=8cm]{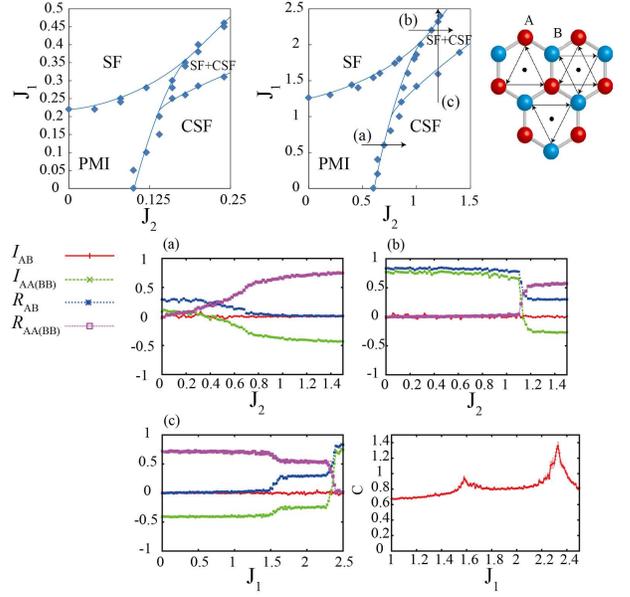}
\caption{Phase diagrams of HBHM with $U=0.1$ (top left) and 
$U=10$ (top middle). 
Calculations of the various physical quantities are shown, which 
are used for identification of the phases.
Between the SF and CSF, there exists a coexisting SF+CSF phase.
``Specific heat " $C$ along the line (c) in the phase diagram (bottom right) exhibits 
two peaks at $J_1\simeq 1.6$ and $2.3$.
%%%
The system size is $L=6$}
\label{Honeycomb_PD}
\end{figure}
%%%%%%%%%%%%%%%%%%%%%%%%%%%%%%%%%%%%%%%%%%%%%%%%%%%%%%%%%%%%%%%%%%%%

For the numerical calculation, we consider the following system;
\begin{enumerate}
\item We consider the unit-filling case and in most of the calculations 
we set the variational parameters $\rho_{i}=\rho=1$ for simplicity.
Full MC simulation of the action (\ref{eff_HBHM}) verifies the validity of the
above assumption.
\item We put $\phi=\pi/2$ as in Ref.\cite{Hofstetter}.
\item Both the torus and cylinder geometries are considered.
See Fig.\ref{honeycomb1}.
\item To classify the physical meaning of the observed states, we measure 
the current operators defined as 
\begin{equation}
I_{ij}\equiv 2 \ \mbox{Im}\Big(J_{ij}\langle e^{i(\theta_i-\theta_j)}\rangle\Big),
\label{current}
\end{equation}
and also the link correlation defined by,
\begin{equation}
R_{ij}\equiv 2 \ \mbox{Re}\Big(J_{ij}\langle e^{i(\theta_i-\theta_j)}\rangle\Big),
\label{R}
\end{equation}
where $J_{ij}$ is $J_1$ and $J_2e^{i\phi}$ for the links connecting NN and NNN sites, respectively.
\end{enumerate}
As explained in the previous work Ref\cite{Hofstetter}, 
for small $J_1/U$ and $J_2/U$, the system is the plaquette Mott insulator (PMI)
that has only finite local correlation $I^{\rm PMI}_{AA(BB)}\neq 0$
and vanishing nonlocal correlations.
On the other hand for a large $J_1\gg J_2$, the ordinary superfluid (SF) forms and 
it has a positive expectation value $I^{\rm SF}_{AA(BB)}>0$ and $I^{\rm SF}_{AB}=0$.
For the weakly interacting limit, $I^{\rm SF}_{AA(BB)}=2\rho J_2$.
For $J_1\ll J_2$, the $J_2$-term in the Hamiltonian Eq.(\ref{HBHM})
dominates and the chiral SF (CSF) forms with 
a negative expectation value $I^{\rm CSF}_{AA(BB)}<0$.
For the weakly interacting limit, $I^{\rm CSF}_{AA(BB)}=-2\rho J_2$.

%%%%%%%%%%%%%%%%%%%%%%%%%%%%%%%%%%%%%%%%%%%%%%%%%%%%%%%%%%%%%%%%%%%%
%Fig.9
\begin{figure}[h]
\centering
\includegraphics[width=9cm]{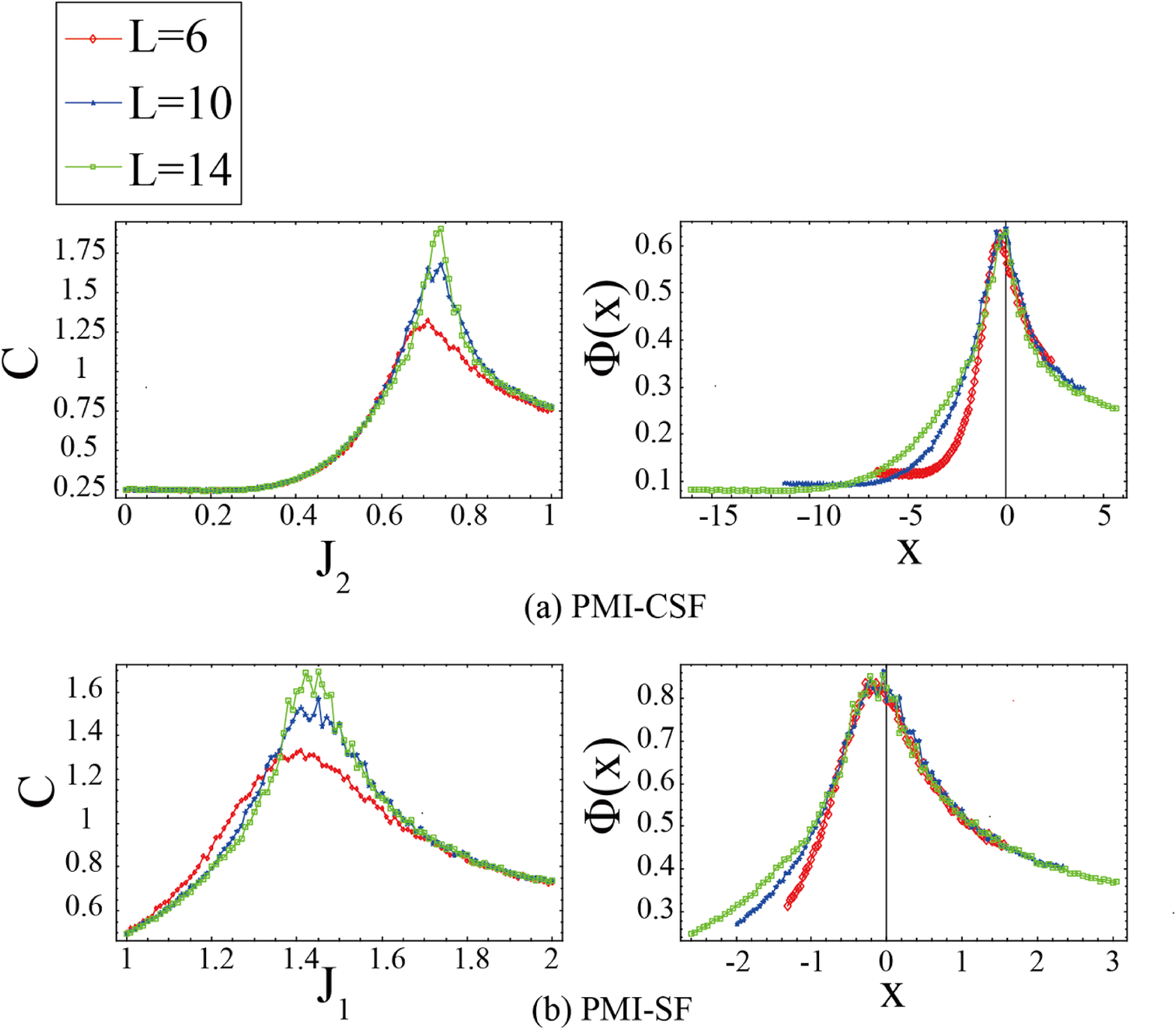}
\caption{``Specific heat" $C$ and the scaling function of the FSS $\Phi(x)$.
Upper panels, the PMI-CSF phase transition.
Lower panels, the PMI-SF phase transition.
FSS indicates that both phase transitions are of second order.
$U=10$.}
\label{FSSMI}
\end{figure}
%%%%%%%%%%%%%%%%%%%%%%%%%%%%%%%%%%%%%%%%%%%%%%%%%%%%%%%%%%%%%%%%%%%%

In this work, the HBHM is studied by the eQMC and the phase boundaries are identified
by calculating the internal energy $E$ and the ``specific heat" $C$ as 
in the study on the BH model in the previous section.
We furthermore calculate the density of states, $N(S)$, to identify the order
of phase transitions, which is defined as 
\begin{eqnarray}
Z_{\rm HBH}&=&\int [d\theta(r)]e^{-S_{\rm HBH}}  \nonumber \\
&=&\int dS \ e^{-S}\int [d\theta(r)] \ \delta(S-S_{\rm HBH}) \nonumber \\
&=&\int dS \ e^{-S}N(S).
\label{N(S)}
\end{eqnarray}
On a phase transition point, we calculate $N(S)$ by the MC simulations.
If $e^{-S}N(S)$ has a double-peak shape as a function of $S$, the phase
transition is of first order, whereas a single-peak shape of $e^{-S}N(S)$ indicates a 
second-order transition.

We first consider the system with the periodic boundary condition, i.e., a torus.
Fig.{\ref{Honeycomb_PD}} exhibits the global phase diagrams of the HBHM
with $U=0.1$ and $10$.
In the phase diagram, there are four phases, three of which are found in 
the previous work\cite{Hofstetter}, i.e., the PMI, SF, and CSF. 
The fourth phase is a coexisting phase possessing both the SF and CSF correlations.
Calculated ``specific heat" $C$ is shown in Fig.\ref{Honeycomb_PD} 
as a function of $J_1$, which indicates the existence of two phase transitions 
for $J_2=1.2$ and $U=10$.
The NN and NNN currents are calculated for the identification of the
phase and the results are shown in Fig.\ref{Honeycomb_PD}.
In the SF, the intra-sublattice current $I_{AA}^{\rm SF}>0$, whereas
in the CSF $I_{AA}^{\rm CSF}<0$ as explained in Ref.\cite{Hofstetter}.
In the SF+CSF coexisting phase, the value of $I_{AA}$, $I^{\rm co}_{AA}$, is 
$I_{AA}^{\rm CSF}<I^{\rm co}_{AA}<I_{AA}^{\rm CSF}$. 
Phase diagram of other values of $U$ has qualitatively the same structure with 
that in Fig.\ref{Honeycomb_PD}.
However, as decreasing on-site interaction $U$, the phase boundary shifts 
because of the suppression of quantum fluctuations of the phase $\theta_{i}$, 
which are induced by growing density fluctuations.  
``Specific heat" $C$ and the scaling function of the FSS are shown in
Fig.\ref{FSSMI}.
The results indicate that both the PMI-SF and PMI-CSF transitions are of second
order. 
%%%
The critical exponents are estimated as
$\nu=0.95 (1.25), \sigma=0.42 (0.26)$ for PMI-SF (PMI-CSF).
%%%
On the other hand for the phase transition from the SF to the SF+CSF phase
and also from the CSF to the SF+CSF phase,
the calculated $C$ for various system sizes does not exhibit the FSS, nor
the density of states $N(S)$ defined by Eq.(\ref{N(S)}) exhibits
a double-peak shape at the phase transition point.
More detailed study using a large-scale systems is needed to clarify the
properties of the phase transition SF (CSF)$\rightarrow$ SF+CSF.

%%%%%%%%%%%%%%%%%%%%%%%%%%%%%%%%%%%%%%%%%%%%%%%%%%%%%%%%%%%%%%%%%%%%
%Fig.10
\begin{figure}[h]
\centering
\includegraphics[width=9cm]{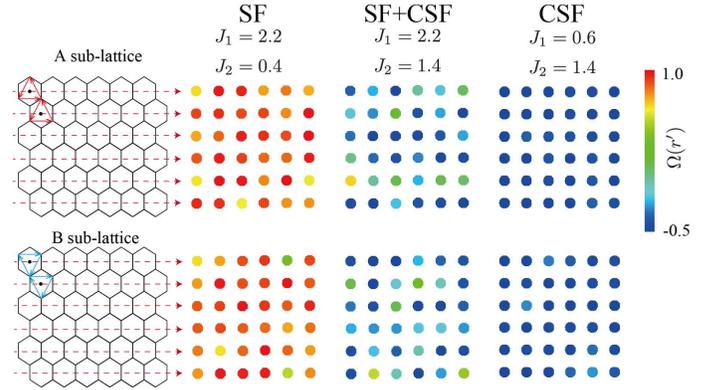}
\caption{Snapshots of vortex $\Omega_A$ and $\Omega_B$ in 
the SF, CSF and SF+CSF phases.
In the SF, phase of BEC has a uniform distribution.
On the other hand in the CSF, $120^o$-structure forms.
In the SF+CSF phase, a spatial mixture of them appears.}
\label{honeycombvortex}
\end{figure}
%%%%%%%%%%%%%%%%%%%%%%%%%%%%%%%%%%%%%%%%%%%%%%%%%%%%%%%%%%%%%%%%%%%%

It is interesting to see if a vortex solid forms in the SF and CSF phases,
and in particular how the vortex solid is deformed by the coexistence
of the SF and CSF in the SF+CSF phase. 
To this end, we calculate the triangle vorticity of the $A$ and $B$-sublattices,
$\Omega_{A(B)}(r)$, which is defined by 
\begin{eqnarray}
\Omega_{A(B)}(r)&=&{1 \over 3}
\Big(\sin(\theta_2-\theta_1+\phi)
+\sin(\theta_3-\theta_2+\phi)  \nonumber \\
&&\hspace{0.5cm} +\sin(\theta_1-\theta_3+\phi)\Big),
\label{trivortex}
\end{eqnarray}
where the detailed definition of $\theta$'s, see Fig.\ref{honeycombvortex} and 
$\phi=\pi/2$ in the
present case.
For the configuration of the uniform $\theta_i(r)$, $\Omega_{A(B)}(r)=1$,
whereas for the $120^o$-configuration, $\Omega_{A(B)}(r)=-1/2$.
Snapshots of $\Omega_{A(B)}(r)$ are shown in Fig.\ref{honeycombvortex}, and 
the results indicate
that the uniform-$\theta$ configurations are realized in the SF, whereas
nearly  $120^o$-configurations are in the CSF, as expected.
On the other hand, the SF+CSF phase has a nontrivial distribution of the vortex.
From this result, we expect that immiscible SF droplets exist in the CSF and 
vice versa in the SF+CSF state..

%%%%%%%%%%%%%%%%%%%%%%%%%%%%%%%%%%%%%%%%%%%%%%%%%%%%%%%%%%%%%%%%%%%%
%Fig.11
\begin{figure}[t]
\centering
\includegraphics[width=8.3cm]{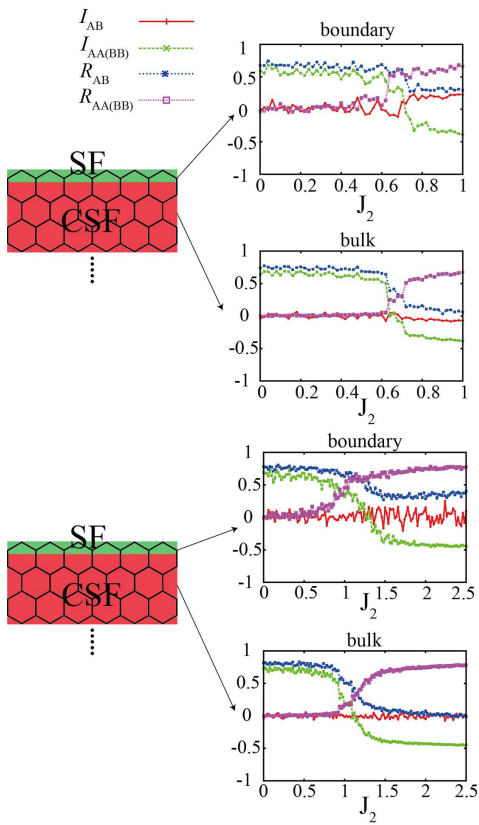}
\caption{Phase diagrams of HBHM in the cylinder geometry. 
Calculations of the various physical quantities are shown, which 
are used for identification of the phases.
Near zigzag edges, the quasi-one-dimensional SF state forms in the
bulk CSF state.
Upper panel for $U=0.5$ and lower panel $U=10$.
%%%
The system size is $L=6$}
\label{edge}
\end{figure}
%%%%%%%%%%%%%%%%%%%%%%%%%%%%%%%%%%%%%%%%%%%%%%%%%%%%%%%%%%%%%%%%%%%%

Finally, let us study the HBHM in the cylinder geometry shown in Fig.\ref{honeycomb1}.
In particular we are interested in BEC behavior near the zigzag edges of the cylinder.
In recent experiments, such a sharp edge boundary can be created by using 
the optical-box trap method\cite{box}.
For the SF phase of the HBHM, Bogoliubov excitations near the edges was
recently studied\cite{furukawa2}.
On the other hand in the present study, we focus on the SF+CSF phase 
that was observed by the eQMC in the previous section.  
It is expected that near the edges a quasi-one-dimensional excitation forms
and the $J_1$-term in Eq.(\ref{HBHM}) dominates there.
Therefore, the SF state appears near the edges of the cylinder.
This expectation is verified by calculating $I_{ij}$ and $R_{ij}$.
See Fig.\ref{edge}.
Schematic picture is that for the bulk CSF state, 
the SF state appears near the edge of the cylinder BEC, whose bulk
state is the CSF for $J_2>J_{2c}$, where $J_{2c}$ is the critical value of $J_2$.
On the other hand for smaller $J_2<J_{2c}$, the whole system is the SF.

%%%%%%%%%%%%%%%%%%%%%%%%%%%%%%%%%%%%%%%%%%%%%%%%%%%%%
\section{Conclusion}

In this paper, we studied the BHM and HBHM with complex hopping amplitudes
that are recently realized by experiments of the cold 
atoms\cite{staggeredmg,uniformmg,Honeycomb_ex}.
We first explained the numerical methods that we call the eQMC.
Most of the numerical results in the present paper have been obtained by
the eQMC.

For the BHM in various magnetic fields, we focused on the vortex-solid state
and clarified its existence for various flux quanta $f=p/q$.
Various spatial patterns of the vortex-solid have been identified by 
calculating the winding number for each lattice plaquette.
We furthermore found that the on-site repulsion $U$ plays an important role
for the phase transition from the vortex solid to vortex liquid.
By the duality transformation of the effective mode of the BHM, 
we discussed the condition on $f$ for the formation of the vortex solid.

For the HBHM, we have obtained the global phase diagram of the ground-state
with the complex NNN hopping with $\phi=\pi/2$.
Besides those found in Ref.\cite{Hofstetter}, we have found another phase in which 
the SF and CSF regions coexist.
Uniform and $120^o$ structures of the phase degrees of the freedom of the
condensed boson field are `entangled' with each other.
More detailed study is needed to clarify the dynamical properties of the
SF+CSF phase like a quantum (in)stability, etc.
This is a future problem.
Finally, we studied the SF+CSF state of the HBHM in the cylinder geometry, and
found that the quasi-1D SF state appears near the edges of the
cylinder.
This phenomenon might have some connection to the edge state in the 
quantum Hall like state.
This is also a future problem.

\acknowledgments
Y. K. acknowledges the support of a Grant-in-Aid for JSPS
Fellows (No. 15J07370). This work was partially supported by Grant-in-Aid
for Scientific Research from Japan Society for the 
Promotion of Science under Grant No.26400246.

%%%%%%%%%%%%%%%%%%%%%%%%%%%%%%%%%%%%%%%%%%%%%%%%%%%%%%%%%%%%%%%%%%%

\end{document}